# Coherent effects contribution to a fast gate fidelity in ion quantum computer

Pavel Sidorov[1], Mikhail Aksenov[1], Ilia Zalivako[1], Alexander Borisenko[1], Ilya Semerikov[1], Ksenia Khabarova[1,2] and Nikolai Kolachevsky[1,2]

[1] *P.N. Lebedev Physical Institute, 53 Leninsky prospekt, 119991, Moscow, Russia*
[2] *Russian Quantum Center, Skolkovo, 143025, Moscow, Russia*

Trapped ions are one of the most promising platforms for quantum computing due to the longest qubit coherence times and the highest gate fidelities. However, scaling the number of ions (qubits) in a linear Coulomb crystal is the key difficulty on the way to multi-qubit systems. One of the promising pathways to scale the number of qubits is to implement the pulsed non-adiabatic gates based on the sequence of State Dependent Kicks (SDKs). We have analytically and numerically studied the influence of coherent effects in the SDK sequence and, correspondingly, have deduced the influence of the individual SDK error on the net gate's fidelity. We have shown that the coherence effects significantly impact the fidelity of non-adiabatic gates and must be taken into the account. As practical examples, we have developed a numerical model for full simulation of coherence effects using a linear ion microtrap array and a 2D microtrap array. We have also studied the dependency of the gate fidelity on the laser power fluctuations.

## I. INTRODUCTION

Quantum computers and simulators are considered as promising systems to solve certain computational tasks much more efficiently than classical computers [1–3]. The basic unit of quantum information (qubit) is regularly represented as a two-level system which can be implemented in practice using superconducting junctions [4], ions [5], quantum dots [6], neutral atoms [7], photon states [8] and many other systems. Today, trapped ions are considered as one of the leading platforms for quantum computation due to the longest qubit coherence times [9], high qubit state preparation and readout fidelities [10], and high fidelity of quantum gates [11]. Ion quantum computers have won leadership in the quantum volume parameter, which jointly characterizes the number of qubits and the fidelity of operations [12]. Still, the acknowledged problems are scaling of the number of ions in linear chains and increasing the gate speed.

Most of the implementations utilize linear ion crystals in a Paul trap [13,14]. The information is encoded in two electronic levels of each ion, while the collective ion motion is used for entanglement. The size of the quantum register based on this architecture demonstrated to-the-date and used for quantum simulations is of 53 qubits [15]. There are two factors which should be considered on the way to increase the number of ions in the 1D register. First, for the most well-known gates like Cirac-Zoller gate [16] or Mølmer–Sørensen gate [17], the ion motional frequencies should be maintained as high as possible, because the execution time for adiabatic gates is much longer than an ion oscillation period. On the other hand, large ion crystals with high axial frequencies tend to change configuration from the linear to the 2D or the 3D structures which prevents individual addressing and distorts the spectrum of motional modes. The requirements are contraversive, and one should find a compromise between axial confinement and the number of ions.

On the way to overcome these difficulties several architectures were proposed. One of the alternatives is to use a number of small quantum registers which can be entangled with each other. The entanglement can be performed using photons [18,19] or by physical transporting of ions carrying quantum information [20]. Another approach is to use several traps with a small trap-to-trap separation providing direct interaction between ions in different registers [21,22]. In both cases one can provide high motional frequencies, relatively simple vibrational spectrum and high-quality addressing.

These approaches lack the important feature of traditional single-register ion quantum computers – direct all-to-all qubit connectivity. Still, the connectivity inside each of the registers is full, while the connectivity between registers can be designed relatively high via thoughtful choice of interconnects topology. For instance, in the case of several closely spaced traps an auxiliary set of single-ion traps can be placed in the nodes of the 2D mesh [23]. In this case each single-qubit register has multiple neighbors available for entangling.

Significant increase of the gate speed can be achieved by implementation of non-adiabatic (fast) quantum gates [24]. The background idea of the fast gates is the same as for the adiabatic gates – electronic states of the ions are entangled via collective motion caused by the Coulomb interaction and external forces (e.g. light fields). The external force used for operation should depend on the ion state. At the end of operation, the electronic and the motional states should be decoupled to prevent decoherence caused by heating.

In the case of adiabatic gates, the external force excites motion on a timescale much larger than a single period of ion secular motion. For non-adiabatic gates, strong forces act on particles for the time interval shorter (or on the order) than the oscillation period. Therefore, mechanical perturbation caused by the external force does not spread far along the system during the operation. It allows to efficiently suppress the undesired influence of distant ions (typically not involved in the operation) and provides high fidelity even in large systems.

Non-adiabatic gates were carefully theoretically and experimentally [25,26] studied since the first proposal [24]. Various pulse optimization schemes were suggested [24,27,28]. The dependency of achievable gate fidelity on different experiment parameters, such as the number of ions, the number of pulses, the trap configurations and the noise sources were analyzed [29–32]. It was shown theoretically, that with the increase of the number of pulses the gate error can be reduced below $10^{-5}$. Moreover, the increase of the number of ions does not cause significant reduction of the fidelity. In [30] it was shown, that with the increase of the number of ions the gate error does not exceed $10^{-5}$ which is an important step towards fault-tolerant scalable ion based quantum computations.

Despite positive theoretical expectations, the experimental realization demonstrated the pulsed non-adiabatic gate fidelity of only 76 % [25]. Several explanations were proposed, including the pulse-picking scheme imperfections and the influence of micromotion.

In this paper we study several factors which can reduce the gate fidelity and cause the discrepancy between the experimental results and existing theoretical predictions. In particular, we study two sources of errors: (i) coming from coherent build-up of single pulses' imperfections and (ii) resulting from the laser power fluctuations. The coherent effects are especially important for the hyperfine qubits (e.g. microwave $^{171}Yb^+$ qubits) where Raman laser pulses are used for the entanglement of particles (see [25]). We also model the situation when the ions are placed not in the single linear trap, but in the neighboring individual microtraps. Finally, we use the developed mathematical model to simulate the gate operation on two neighboring ions in the 2D microtrap array.

In the section II we briefly describe the basics of pulsed non-adiabatic gates. Section III contains analytical treatment of the coherent build-up of individual pulse errors. In the section IV we perform full numerical simulation of the two-qubit gates (including coherence effects) in the different trap configurations. Section V is devoted to the effect of laser power fluctuations. This is followed by the discussions in the section VI.

## II. PULSED NON-ADIABATIC QUANTUM GATES

Pulsed non-adiabatic quantum gates are based on the sequence of spin-dependent momentum kicks (SDKs) arising from an ion interaction with a laser pulse [28]. Depending on the electronic quantum state of the ion, $|0\rangle$ or $|1\rangle$, SDK brings ion to motion in different directions. The duration of SDK is assumed to be much shorter than the oscillation period of the ion in the trap. The evolution operator of the SDK performed simultaneously on two ions can be written in the following form [33]:

$$U_{kick} = e^{-2i\vec{k}(\vec{r_1}\sigma_1^z + \vec{r_2}\sigma_2^z)}, \quad (1)$$

where $2\vec{k} = \frac{\overline{\Delta p}}{\hbar}$, $\overline{\Delta p}$ is the momentum transferred to the ions, $\vec{r_1}, \vec{r_2}$ - position operators of the ions, $\sigma_1^z, \sigma_2^z$ are the Pauli operators acting on the electronic states of the corresponding ions. Expressing ions motion as a superposition of normal oscillation modes, we can rewrite (1) in the form of displacement operators $D_p$ for the mode $p$:

$$U_{kick} = \prod_{p=1}^{3L} \widehat{D}_p(-2i(\vec{b}_1^{(p)}\sigma_1^z + \vec{b}_2^{(p)}\sigma_2^z)\vec{\eta}_p), \quad (2)$$

where $\vec{\eta}_p = \sqrt{\frac{\hbar}{2m\omega_p}}\vec{k}$ is the Lamb-Dicke parameter and $\vec{b}_i^{(p)}$ – the oscillation amplitude of the i-th ion for the p-th mode and L is number of interacting ions.

It follows that the most natural way to analyze the evolution of the mode states is to represent them in phase space in the rotating frame, where the coherent state $|\alpha_p\rangle$ is taken as an initial state. Further we will show that selecting a certain initial state causes no loss of generality. According to (2), each SDK will lead to the motional state displacement and acquiring a phase called a geometric phase. If SDKs are repeated one after another much faster than the period of ion oscillations for z times, then the displacement of the coherent state will be repeated z times in the same direction on the phase plane. If there is an interval $t$ between several subsequent SDKs, then each next pulse will shift the state in the direction rotated by an angle $\omega_p t$ with respect to the previous SDK due to the free evolution of the coherent state. An example of a phase-space trajectory for the vibrational mode state after an SDK sequence is shown in Fig. 1.

If the phase space trajectory closes at the end of the SDK sequence, the mode state evolves as $|\alpha_p\rangle \rightarrow e^{i\xi_p}|\alpha_p e^{-i\omega_p T}\rangle$, where $\xi_p$ is the total geometric phase. The

latter is proportional to the area enclosed by the trajectory in the rotating phase space [34].

Here $T$ is the total duration of the sequence. It can be shown that the sign of the acquired geometric phase $\xi_p$ is proportional to the parity of the internal states of the entangled ions $\sigma_1^z \sigma_2^z$.

Hence, by proper adjustment of the number of kicks and the arrival times providing that the trajectories in the phase space for all modes involved are closed, one performs high-quality two-qubit gate. At the end of operation, the internal state is disentangled from the vibrational state and the truth table of the operation is:

$$|00\rangle \to e^{i\Phi}|00\rangle$$
$$|01\rangle \to e^{-i\Phi}|01\rangle$$
$$|10\rangle \to e^{-i\Phi}|10\rangle$$
$$|11\rangle \to e^{i\Phi}|11\rangle,$$

where
$\Phi = \sum_{p=1}^{3L} \xi_p = 8\sum_{p=1}^{3L}(\vec{\eta}_p \vec{b}_1^{(p)})(\vec{\eta}_p \vec{b}_2^{(p)}) \sum_{m=2}^{N} \sum_{k=1}^{m-1} z_m z_k \sin(\nu_p(t_m - t_k))$. (3)

To maximize entanglement, the phase of $\Phi=\pi/4$ is typically selected. Thus, one needs to find the pulse sequence providing the conditions

$$\sum_{k=1}^{N} z_k e^{i\omega_p t_k} = 0, \quad (4)$$

$$\Phi = \pi/4. \quad (5)$$

The condition (4) provides the closure of the corresponding phase-space trajectories for all the vibrational modes. This also explains why consideration of coherent state as the initial state does not result in the loss of generality: any arbitrary initial state can be decomposed over complete basis of coherent states, any motional state will be restored at the end of the operation. One cannot strictly fulfil both conditions (4) and (5) because the number of independent parameters is limited. It causes finite theoretical fidelity of the operation. We define fidelity of the operation as $F = |\langle U\psi_0|\psi_{id}\rangle|^2$, where U – gate evolution operator, $|\psi_0\rangle = (\frac{1}{2}|00\rangle + \frac{1}{2}|01\rangle + \frac{1}{2}|10\rangle + \frac{1}{2}|11\rangle) \otimes \prod_{p=1}^{3L}|0\rangle_p$, $|\psi_{id}\rangle = \frac{1}{2}i|00\rangle + \frac{1}{2}|01\rangle + \frac{1}{2}|10\rangle + \frac{1}{2}i|11\rangle \otimes \prod_{p=1}^{3L}|0\rangle_p$. In the case of the non-adiabatic pulsed gate an analytical expression can be obtained:

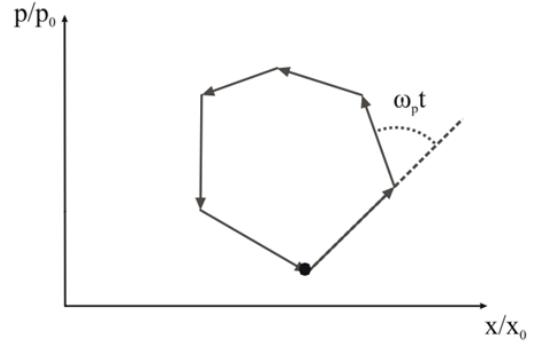

FIG. 1. Evolution of the normal mode coherent state in the rotating phase space under the action of an SDK sequence. The bold point marks the initial coherent state. Here $p_0 = \sqrt{\frac{\hbar m \omega_p}{2}}$, $x_0 = \sqrt{\frac{\hbar}{2m\omega_p}}$.

$$F = \frac{1}{16}\left[4 + 2\exp\left(\sum_{p=1}^{3L} -4\left((\vec{b}_1^{(p)} - \vec{b}_2^{(p)})\vec{c}_p\right)^2 \left(\frac{1}{2} + \bar{n}_p\right)\right) + 2\exp\left(\sum_{p=1}^{3L} -4\left((\vec{b}_1^{(p)} + \vec{b}_2^{(p)})\vec{c}_p\right)^2 \left(\frac{1}{2} + \bar{n}_p\right)\right) + 4\left(\exp\left(\sum_{p=1}^{3L} -4\left(\vec{b}_1^{(p)}\vec{c}_p\right)^2 \left(\frac{1}{2} + \bar{n}_p\right)\right) + \exp\left(\sum_{p=1}^{3L} -4\left(\vec{b}_2^{(p)}\vec{c}_p\right)^2 \left(\frac{1}{2} + \bar{n}_p\right)\right)\right)\sin(2\Phi)\right], \quad (6)$$

$\bar{n}_p$ is the average phonon population of the $p$-th motional mode, $\vec{c}_p = 2\vec{\eta}_p \sum_{k=1}^{N} z_k e^{i\omega_p t_k}$. In practice, the maximal fidelity is lower than (6) due to various technical factors.

All the statements given above are applicable to any pulsed non-adiabatic gates. Depending on the qubit type (microwave or optical) the way how the operator (1) is implemented differs. In the case of the optical qubit, operator (1) can be obtained by applying two successive counter-propagating laser π-pulses resonant to the qubit transition. In the case of the microwave qubit, Raman pulses are used. The evolution operator in this case becomes more complicated due to the Kapitza-Dirac scattering of the atomic wave function on the standing light wave formed by two laser pulses [35].

If two Raman pulses with the duration of much less than the ion oscillation period is applied, the interaction evolution operator is given by [36]:

$$U_{t,j} = \sum_{n=-\infty}^{\infty} i^n J_n(\theta) e^{in\Delta\phi(t)} \sigma_{x_j}^n \prod_{p=1}^{3L} D_p\left[in\vec{b}_j^p \vec{\eta}_p\right], \quad (7)$$

where t- is the arrival time of the pulses, j – the number of the ions, Δϕ(t) – total optical phase difference between two pulses, θ – the Raman pulse area. The pulses can be derived, for example, by splitting a frequency comb laser beam and shifting the frequency of one of the beams by $\omega_{AOM}$ with an acousto-optical modulator in such a way, that each comb tooth has a counterpart shifted by the qubit frequency $\omega_{qubit} = \pm\omega_{AOM} + nfrep$. Operators similar to (2) are obtained, while high diffraction orders are suppressed by composing SDK of several successive laser pulses and choosing optimal delays between them:

$$U_{SDK,pos} = ie^{i\phi_0}\prod_{p=1}^{3L} D_p\left[ i\vec{b_j^p}\vec{\eta_p}\right]\hat{\sigma}_{+,j} \\ + ie^{-i\phi_0}\prod_{p=1}^{3L} D_p\left[-i\vec{b_j^p}\vec{\eta_p}\right]\hat{\sigma}_{-,j}, \quad (8)$$

$$U_{SDK,neg} = ie^{-i\phi_1}\prod_{p=1}^{3L} D_p\left[ i\vec{b_j^p}\vec{\eta_p}\right]\hat{\sigma}_{-,j} \\ + ie^{i\phi_1}\prod_{p=1}^{3L} D_p\left[-i\vec{b_j^p}\vec{\eta_p}\right]\hat{\sigma}_{+,j}. \quad (9)$$

Here $\phi_0$, $\phi_1$ are the time-independent parts of the phase difference of the optical Raman pulses. Further we will call the operator (8) "the positive SDK" and operator (9) – "the negative SDK". If we apply two such sequences of pulses corresponding to operators (8) and (9) one after another, we get an operator proportional to

$$U_{Total} = e^{i(\phi_0-\phi_1)\sigma_{z,j}}\prod_{p=1}^{3L} D_p\left[2i\vec{b_j^p}\vec{\eta_p}\sigma_{z,j}\right]. \quad (10)$$

For two ions this operator has the same form as the SDK operator (2) up to the phase factor. It is important to note some features of (7): using the same sequence of pulses with the opposite sign of $\Delta\phi(t)$ (which, for example, can be implemented by swapping the directions of laser pulses), operator (8) will turn into (9) and vice versa. Thereby, for operator (10), the direction of the transmitted momentum will be reversed.

### III. COHERENT SDK ERROR SUMMATION

We see from the previous section, that there are several sources of the gate errors. The first one is the imperfect closure of the state trajectories in the motional phase space (given by (6)), while the second is due to non-ideal spin-flips in the qubit state space during each individual SDK. The first type of errors is treated in literature analytically in the fully coherent way. Till now, the second type of errors were considered uncorrelated for each SDK. It means, that the phase of any π-pulse error is considered to be random for each of the SDKs.

Here we study the impact of coherence between individual SDK errors on the gate fidelity both analytically and numerically. In particular, we make our analysis for microwave qubits excited by Raman pulses.

First of all, we qualitatively demonstrate the effect by considering two successive SDKs: positive and negative coming one after another. In the ideal case it should give us the evolution operator (10).

In the interaction picture exact evolution operator corresponding to the positive SDK takes the form of [37]:

$$U_{pos} = 1 + \frac{i\Theta}{2N}\left\{\left[e^{i\phi_0}D[i\eta]\left(\sum_{k=1}^{N} e^{iq_+t_k}\right)\hat{\sigma}_+ \\ + e^{-i\phi_0}D[-i\eta]\left(\sum_{k=1}^{N} e^{-iq_+t_k}\right)\hat{\sigma}_-\right] \\ + \left[e^{i\phi_0}D[i\eta]\left(\sum_{k=1}^{N} e^{iq_-t_k}\right)\hat{\sigma}_- \\ + e^{-i\phi_0}D[-i\eta]\left(\sum_{k=1}^{N} e^{-iq_-t_k}\right)\hat{\sigma}_+\right]\right\} \\ + O((\Theta/N)^2). \quad (11)$$

Here $q_\pm = \omega_{qubit} \pm \omega_{AOM}$. When $t_k$ obey the resonance condition, $(\sum_{k=1}^{N} e^{iq_+t_k}) = N$ and $(\sum_{k=1}^{N} e^{iq_-t_k}) = C$, $|C| \ll N$, the final expression becomes close to the (8). Expression for negative SDK $U_{neg}$ can be obtained by replacing $q_+ \leftrightarrow q_-, \phi_0 \to \phi_1$ in (11). The resulting evolution operator for the pair of such pulses will be just their product, as they are written in the interaction picture. The resonant terms of the operator will be proportional to $\hat{\sigma}_+\hat{\sigma}_-D[2i\eta]$ and $\hat{\sigma}_-\hat{\sigma}_+D[-2i\eta]$ and give us (10). Although, one can notice that there are other parasitic terms on the order of $O(C/N)$, which cause a deviation from the (10) and result in a gate errors:

$$\left[\frac{C^*}{N}e^{i(\phi_0-\phi_1)-iq_-t_{N+1}} + \frac{C^*}{N}e^{i(\phi_1-\phi_0)+iq_+t_{N+1}}\right]\hat{\sigma}_+\hat{\sigma}_- + H.c. = \\ \frac{2C^*}{N}e^{i\omega_{AOM}t_{N+1}}cos(\phi_0 - \phi_1 + \omega_{qubit}t_{N+1})\hat{\sigma}_+\hat{\sigma}_- + H.c. \quad (12)$$

Here $t_{N+1}$ is the arrival time of the first pulse of the negative SDK. This expression shows that if the time interval between positive and negative SDKs changes, the absolute value of (12) undergoes oscillations with the qubit frequency $\omega_{qubit}$. In the experiment SDKs are usually separated by an integer number of the laser repetition periods so the gate error described by these terms will be proportional to the $\cos(2\pi T_{rep}/T_{qubit})$. Due to the same reason such oscillatory behavior is expected to appear also in the net gate fidelity. The amplitude of these oscillations according to (12) is proportional to the individual SDK error. It means that with the increase of the number of pulses in a single SDK and consequent improve in its fidelity, the sensitivity of the gate fidelity to the repetition rate will decrease.

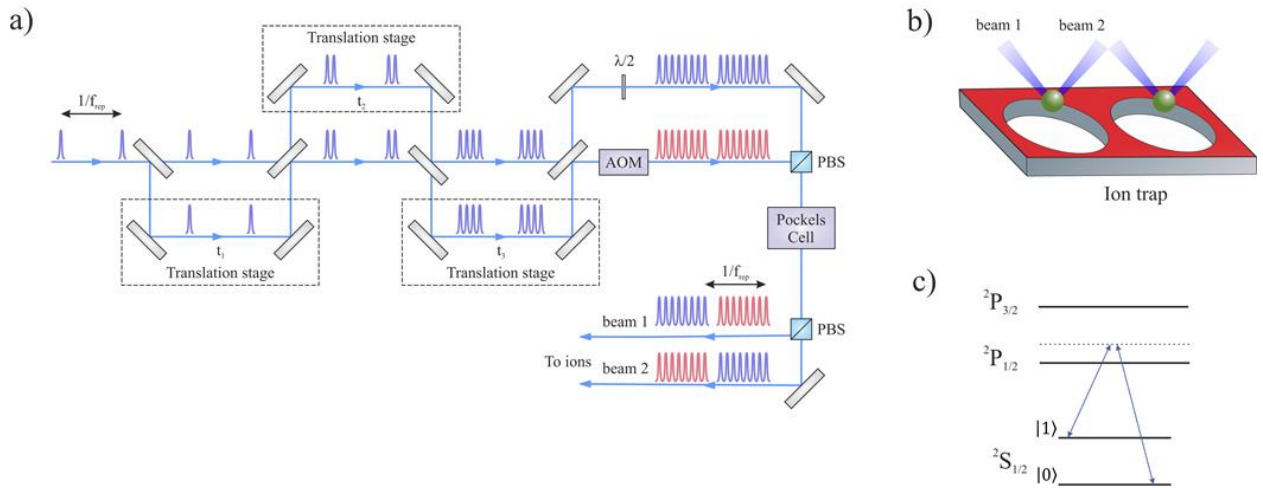

FIG. 2. (a) Proposed experimental design for the implementation of SDK. PBS – polarizing beam splitter. Acousto-optic modulator controls relative phase and frequency of pulse trains, while the Pockels cell allows to send corresponding train to the certain beam path. (b) Ions addressing in ion traps. (c) Simplified Raman transition scheme in case of $^{171}$Yb$^+$ ion.

This treatment, however, is valid only when a small number of pulses in SDK is considered. For larger number of the pulses, the first order parasitic terms in the expansion will be cancelled and higher order terms will become more and more important. The dependency on the $T_{rep}$ will be changed, although the qualitative picture remains valid.

It is important to not confuse this effect with the gate fidelity dependence on the repetition rate, which is shown e.g. in [31]. In our work the effect comes from the individual SDK errors and the way how they sum up which depends on the $T_{rep}$. In [30] the error is caused by non-ideal motional trajectories closure, so the dependency is caused by the optimization problem restrictions (such errors will be present even in the case of ideal SDKs).

To validate our approach and to take into the account high-order terms of (11) we performed a full numerical simulation of the sequences of SDKs taking into account coherent effects.

## IV. NUMERICAL GATE SIMULATION INCLUDING COHERENT EFFECTS

Parameters of the optimal pulses sequence forming the SDK and the SDKs themselves depend on the particular experimental setup (the trap parameters and the ion). For our analysis we use the setup close to the one described in a [25]. It is also similar to the setup in our laboratory which we plan to use for experimental study of fast gates. The optimization of the gate fidelity (excluding the coherent error summation) is extensively studied by other groups [29,33,38].

In this work, we optimized parameters of the SDK for $^{171}$Yb$^+$ ion in the surface point trap with a radial frequency of 33.5 kHz which is close to our experimental parameters. By choosing the SDK and the gate parameters we keep in mind an experimental scheme shown in Fig. 2. We give its detailed description, which determines the search space for optimizations. Of course, by choosing other experimental scheme (e.g. increasing the number of pulses) it is possible to achieve higher fidelities, but we focus here on the demonstration of coherence effects. We can readily extend our analysis on more complicated configurations, if necessary.

The qubit is encoded in the states $^2S_{1/2}|F = 1; m_F = 0\rangle$ and $^2S_{1/2}|F = 0; m_F = 0\rangle$ of the ion. Raman transitions are driven with a Coherent Paladin Compact laser. The laser emits pulses with duration of about 20 ps at the wavelength λ=355 nm with a repetition rate of $f_{rep} = 120 \pm 2$ MHz and average optical power of P = 4W. Using beamsplitters and delay stages each laser pulse is split into a series of pulses, forming an SDK. In the end the beam is split into two parts and focused onto the ions from different directions in such a way that pulses from different paths temporally overlap on the ions. An acousto-optic modulator (AOM) in one of the paths allows to precisely control phase and frequency shifts. A Pockels cell installed before the last polarization beam splitter allows to choose from which side the AOM-modulated pulses will hit the ions and, therefore, will positive or negative SDK be implemented.

The elaborated approach to SDK optimization can be found in [36]. It can be seen that with increase in number of pulses inside SDK its fidelity gets higher, although it makes both experimental realization and full simulation more difficult, so we chose number of pulses to be 8, as this is the minimal pulse number after which a single SDK fidelity is above 99%.

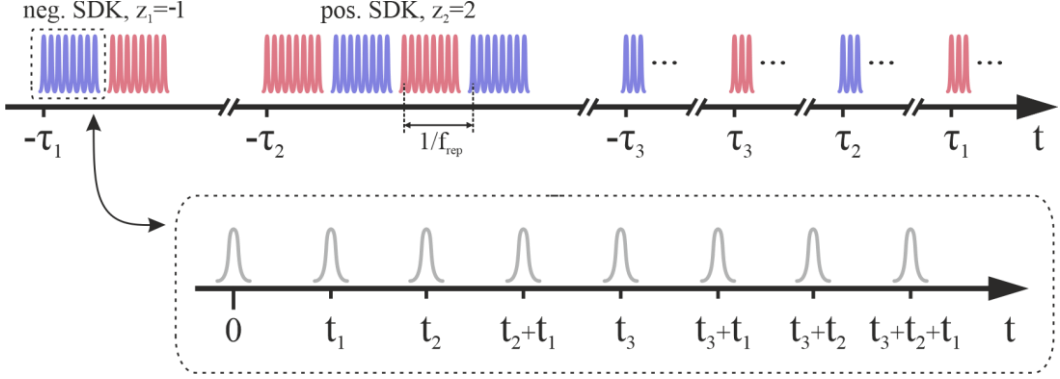

FIG. 3. FRAG strategy with n=1 for proposed scheme. Pulses coming to the ions from one side are shown. Pulses forming positive SDK are colored red, pulses forming negative SDK are colored blue.

The delays $t_1, t_2, t_3$ giving arrival times of the pulses inside the SDK in form $0, t_1, t_2, t_1 + t_2, t_3, t_3 + t_1, t_3 + t_2, t_1 + t_2 + t_3$ (Fig. 3) were obtained by the global optimization under the condition that pulses interact with ions at times $\tilde{t}_k = m/(f_{qubit} + f_{AOM})$. The number $m$ and the AOM frequency $f_{AOM}$ (the initial frequency of 500 MHz) were varied. By optimization we achieved the maximal fidelity of 99.9% for the SDK operator with 8 pulses. The residual error is associated with the incomplete suppression of the higher diffraction orders. The average laser power required to implement this scheme with Raman beams focused to a beam waist radius of 5 μm is close to 4 W.

To find SDK time ticks we chose the FRAG strategy [33], since it allows the fastest geometrical phase acquiring and requires less SDKs. It implies the following SDKs sequence:

$$z_k = \{-n, 2n, -2n, 2n, -2n, n\} (n \in N), \tau_k = \{-\tau_1, -\tau_2, -\tau_3, \tau_3, \tau_2, \tau_1\},$$

where $z_k$ corresponds to the number and the sign of SDKs applied and $\tau_k$ corresponds to the time mark (Fig. 3). Obviously, the difference between different $\tau_k$ should be a multiple to the laser repetition periods $1/f_{rep}$.

We use the least possible number of SDKs with $n = 1$ to reduce effects of single SDK error. Minimal number of pulses also simplifies the full optimization. The times $\tau_1, \tau_2, \tau_3$ (varied in the range of 0 to 100 μs) were found by the global optimization method in order to maximize the total fidelity of the gate according to equation (6). Note, that in (6) all the SDKs providing evolution with operator (4) were assumed to be ideal.

We carry out full numerical simulation to study the contribution of individual SDKs errors and their interference.

Diffraction up to $N = 4$ order is considered. Each laser pulse inside SDK corresponds to the following evolution operator:

$$U = \sum_{n_2=-N}^{N} \sum_{n_1=-N}^{N} (i^{n_1+n_2} J_{n_1}(\theta) J_{n_2}(\theta) e^{i(n_1+n_2)\Delta\phi(t_0)} \cdot \hat{\sigma}_{x,1}^{n_1} \sigma_{x,2}^{n_2} \prod_{p=1}^{3L} D[in_1\vec{b}_1^{(p)}\vec{\eta}_p + in_2\vec{b}_2^{(p)}\vec{\eta}_p]) \quad (13)$$

The initial state diffracts and transforms into a superposition of 2N+1 states. The evolution of vibrational states between the laser pulses constituting the single SDK are neglected. The duration of a single SDK is 3 ns which is indeed much smaller than the oscillation period of the fastest mode. The evolution of quantum states between the SDKs is described by the operator:

$$U_{FE} = e^{-i\omega_{qubit} T(\hat{\sigma}_{z,1}+\hat{\sigma}_{z,2})/2} \prod_{p=1}^{3L} e^{-i\omega_p T a^+ a}, \quad (14)$$

where T – is the time interval between the SDKs and L – number of interacting ions. States with probability amplitudes less than 10-4 were neglected, a discretization step of the coherent vibrational states was set to $|\Delta\alpha|$=10-5. The state $|\psi_0\rangle = \left(\frac{1}{2}|00\rangle + \frac{1}{2}|01\rangle + \frac{1}{2}|10\rangle + \frac{1}{2}|11\rangle\right) \otimes \prod_{p=1}^{3L}|0\rangle_p$ was taken as the initial state. The resulting fidelity was determined by the expression $F = |\langle U\psi_0|\psi_{id}\rangle|^2$, where $|\psi_{id}\rangle = \frac{1}{2}ie^{i\gamma}|00\rangle + \frac{1}{2}|01\rangle + \frac{1}{2}|10\rangle + \frac{1}{2}ie^{-i\gamma}|11\rangle \otimes \prod_{p=1}^{3L}|0\rangle_p$.

In addition to the geometrical phase there is an extra phase $\gamma$ due to the phase factor in operator (10). Its value can be simply calculated analytically, but in our case, it was also found by numerically by maximizing the fidelity.

Fig. 4 shows the dependency of the four successive SDKs (positive, negative, positive, negative) fidelity $F$ on the repetition rate of the laser. For this simulation, a single ion was stored in a surface point trap with a radial frequency of 33.5 kHz. The oscillations qualitatively predicted in the previous section can be observed readily. The oscillation period corresponds the predicted above, while the envelope shape is determined by the number and timing of SDKs and the higher-order SDK errors.

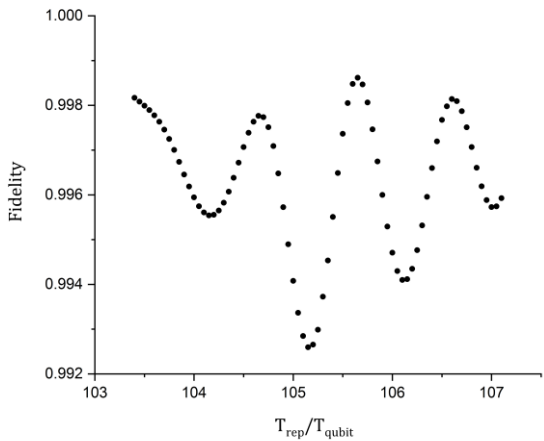

FIG. 4. Fidelity of the sequence of four successive SDKs depending on a repetition period of the laser.

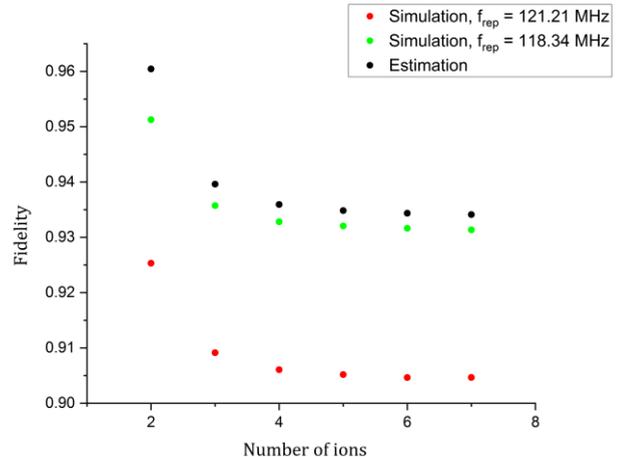

FIG. 5. Two-qubit gate fidelity depending on the size of the quantum register (Number of ions) at different laser repetition rate $f_{rep}$. Black circles show estimation based on the assumption (6) of the uncorrelated SDK errors.

Further we investigate how this effect impacts the quality of quantum operations by increasing the number of ions using fast gates and microtrap arrays. We simulate two-qubit gates between two neighbouring ions in linear arrays of different number of microtraps identical to the considered above (Fig. 2b). Simulations are performed for two different laser repetition frequencies (Fig. 5). The pulse sequence was optimized once for two neighboring microtraps and was not changed afterwards with the register size increase. The full gate time corresponds to 1.6 trap secular periods. In Fig. 5 we also provide an estimation of the expected fidelity using (6) under assumption of uncorrelated SDK errors. In this case the full fidelity would be a product of a fidelity given by (6) and a single SDK fidelity to the power of number of SDKs in the gate.

One can conclude, that due to coherent summation of the single SDKs, errors of the quantum gates can substantially deviate from the estimations based on the assumption of the uncorrelated SDK errors. We show that the amplitude of the error oscillations depending on the repetition rate is of the same order than the mean gate error. The effect must be taken into account when designing high-efficiency quantum gates. Our simulation also supports previous results obtained by other researchers that with increasing number of ions in the quantum register, the two-qubit gate fidelity rapidly settles to the constant value and does not fall further. It is important for the ion quantum computation scaling.

We also apply developed above model to study scaling of quantum computers with 2D arrays of surface microtraps. In particular, we simulated a two-qubit gate between a pair of ions in a 3×3 traps array (Fig. 6). Each ion is trapped above the individual hole in a dielectric plate coated by metal from both sides. The upper surface is under RF-potential while the bottom surface is grounded.

The laser beams were considered to be at 45 degrees to the plane of the trap with $\overrightarrow{\Delta k}$ directed along the line, connecting two ions (Fig. 2b). Such beams directions provide the fastest phase acquiring as cause ions interaction energy changes the most significantly.

The full simulation at $f_{rep} = 121.21$ MHz resulted in the overall fidelity equal to 91.72%, while the estimation with the assumption of the uncorrelated SDKs is equal to 93.72%. Here we also see that coherent effects significantly influence the gate fidelity.

## V. INFLUENCE OF LASER POWER FLUCTUATIONS

One of the important technical aspects that can affect the fidelity of the operation in the experiment is the intensity fluctuation of the laser source. Here we evaluate the effect of the laser intensity fluctuation on the overall fidelity for the experimental scheme described above.

In this experimental scheme effective Rabi frequency determining Rabi oscillations between qubit levels is proportional to the laser power [36]. Laser power fluctuations or deviation from the optimal value lead to the change of the effective Rabi frequency, and, consequently, to a deviation of $\theta$ from $\pi/8$ in formula (7). Figure 7 shows the fidelity of the considered quantum operation plotted against the deviation of the laser source power from the optimal value.

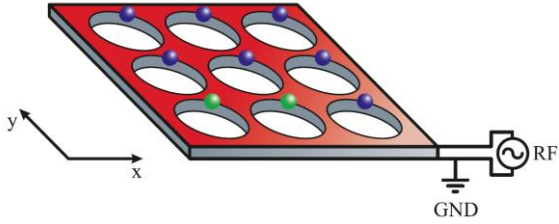

FIG. 6. 3×3 2D point array of ion traps. RF potential is applied to the top surface. The bottom surface is grounded. The intermediate layer consists of dielectric. We perform optimization of entanglement for two of nine ions depicted as green.

This result shows that the gate fidelity is very sensitive to the laser fluctuations. Even 1% power fluctuations can lead to the gate fidelities below 90%. To avoid that careful laser power stabilization is required.

## VI. SUMMARY

In this paper we studied an influence of the coherent summation of the single SDK errors on the fidelity of the non-adiabatic pulsed gates. This effect can lead to the oscillations of the overall gate fidelity depending on the laser repetition rate used for its implementations. The amplitude of these oscillations is of the same order as the mean gate fidelity. This effect, however, does not limit the possibility of scaling ion quantum computers using fast gates, as the fidelity oscillations amplitude decreases with the single SDK error. Thus, improving single SDKs fidelity will lead to significant reduction of this factor.

We have also performed a full simulation (including coherence effects) of the two-qubit gate fidelity depending on the size of the quantum register for the linear microtrap array and a the 2D array. In both cases coherence effects significantly contribute to the gate fidelity. Still, the increase of the quantum register size does not significantly impact the fidelity. It rapidly (after 5 ions in the chain) reaches the constant value and does not significantly change further. It is one of the important advantages on the way to the multi-ion quantum computers.

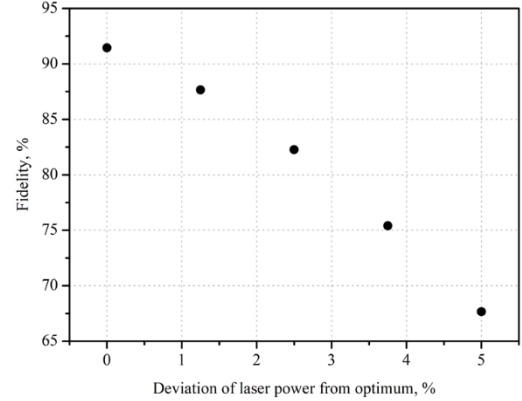

FIG. 7. Operation fidelity versus the deviation of the laser power from the optimal value.

We discussed power fluctuations influence on the two-qubit gate fidelity and showed that high efficiency laser power stabilization is required to perform high-fidelity gates.

## ACKNOWLEDGEMENTS

This work was supported by the Russian Science Foundation (Grant No. 19-12-00274).

[1] P. W. Shor, *M*, in *Proceedings of the 35th Annual Symposium on Foundations of Computer Science* (1994), pp. 124–134.

[2] L. K. Grover, *No Title*, in *Proceedings of the Twenty-Eighth Annual ACM Symposium on Theory of Computing* (1996), pp. 212–19.

[3] Y. Nam, J. S. Chen, N. C. Pisenti, K. Wright, C. Delaney, D. Maslov, K. R. Brown, S. Allen, J. M. Amini, J. Apisdorf, K. M. Beck, A. Blinov, V. Chaplin, M. Chmielewski, C. Collins, S. Debnath, K. M. Hudek, A. M. Ducore, M. Keesan, S. M. Kreikemeier, J. Mizrahi, P. Solomon, M. Williams, J. D. Wong-Campos, D. Moehring, C. Monroe, and J. Kim, *Ground-State Energy Estimation of the Water Molecule on a Trapped-Ion Quantum Computer*, Npj Quantum Inf. **6**, 1 (2020).

[4] G. J. Mooney, C. D. Hill, and L. C. L. Hollenberg, *Entanglement in a 20-Qubit Superconducting Quantum Computer*, Sci. Rep. 1 (2019).

[5] C. D. Bruzewicz, J. Chiaverini, R. McConnell, and J. M. Sage, *Trapped-Ion Quantum Computing: Progress and Challenges*, Appl. Phys. Rev. **6**, (2019).

[6] D. Press, T. D. Ladd, B. Zhang, and Y. Yamamoto, *Complete Quantum Control of a Single Quantum Dot Spin Using Ultrafast Optical Pulses*, Nature **456**, 218 (2008).

[7] M. Saffman, *Quantum Computing with Atomic Qubits and Rydberg Interactions: Progress and Challenges*, J. Phys. B At. Mol. Opt. Phys. **49**, (2016).

[8] J. C. F. Matthews, A. Politi, A. Stefanov, and J. L. O'Brien, *Manipulation of Multiphoton Entanglement in Waveguide Quantum Circuits*, Nat. Photonics **3**, 346 (2009).

[9] Y. Wang, M. Um, J. Zhang, S. An, M. Lyu, J. N. Zhang, L. M. Duan, D. Yum, and K. Kim, *Single-

*Qubit Quantum Memory Exceeding Ten-Minute Coherence Time*, Nat. Photonics **11**, 646 (2017).

[10] T. P. Harty, D. T. C. Allcock, C. J. Ballance, L. Guidoni, H. A. Janacek, N. M. Linke, D. N. Stacey, and D. M. Lucas, *High-Fidelity Preparation, Gates, Memory, and Readout of a Trapped-Ion Quantum Bit*, Phys. Rev. Lett. **113**, 2 (2014).

[11] J. P. Gaebler, T. R. Tan, Y. Lin, Y. Wan, R. Bowler, A. C. Keith, S. Glancy, K. Coakley, E. Knill, D. Leibfried, and D. J. Wineland, *High-Fidelity Universal Gate Set for Be 9 + Ion Qubits*, Phys. Rev. Lett. **117**, 1 (2016).

[12] L. S. Bishop, S. Bravyi, A. Cross, J. M. Gambetta, and J. Smolin, Quantum Volume, 2017.

[13] K. Wright, K. M. Beck, S. Debnath, J. M. Amini, Y. Nam, N. Grzesiak, J.-S. Chen, N. C. Pisenti, M. Chmielewski, C. Collins, K. M. Hudek, J. Mizrahi, J. D. Wong-Campos, S. Allen, J. Apisdorf, P. Solomon, M. Williams, A. M. Ducore, A. Blinov, S. M. Kreikemeier, V. Chaplin, M. Keesan, C. Monroe, and J. Kim, *Benchmarking an 11-Qubit Quantum Computer*, Nat. Commun. 1 (2019).

[14] K. A. Landsman, Y. Wu, P. H. Leung, D. Zhu, N. M. Linke, K. R. Brown, L. Duan, and C. Monroe, *Two-Qubit Entangling Gates within Arbitrarily Long Chains of Trapped Ions*, Phys. Rev. A **100**, 1 (2019).

[15] J. Zhang, G. Pagano, P. W. Hess, A. Kyprianidis, P. Becker, H. Kaplan, A. V. Gorshkov, Z. X. Gong, and C. Monroe, *Observation of a Many-Body Dynamical Phase Transition with a 53-Qubit Quantum Simulator*, Nature **551**, 601 (2017).

[16] F. Schmidt-Kaler, H. Häffner, M. Riebe, S. Gulde, G. P. T. Lancaster, T. Deuschle, C. Becher, C. F. Roos, J. Eschner, and R. Blatt, *Realization of the Cirac-Zoller Controlled-NOT Quantum Gate*, Nature **422**, 408 (2003).

[17] K. Mølmer and A. Sørensen, *Multiparticle Entanglement of Hot Trapped Ions*, Phys. Rev. Lett. **82**, 1835 (1999).

[18] D. Hucul, I. V. Inlek, G. Vittorini, C. Crocker, S. Debnath, S. M. Clark, and C. Monroe, *Modular Entanglement of Atomic Qubits Using Photons and Phonons*, Nat. Phys. **11**, 37 (2015).

[19] V. Krutyanskiy, M. Meraner, J. Schupp, V. Krcmarsky, H. Hainzer, and B. P. Lanyon, *Light-Matter Entanglement over 50 Km of Optical Fibre*, Npj Quantum Inf. **5**, (2019).

[20] J. M. Amini, H. Uys, J. H. Wesenberg, S. Seidelin, J. Britton, J. J. Bollinger, D. Leibfried, C. Ospelkaus, A. P. Vandevender, and D. J. Wineland, *Toward Scalable Ion Traps for Quantum Information Processing*, New J. Phys. **12**, (2010).

[21] J. I. Cirac and P. Zoller, *A Scalable Quantum Computer with Ions in an Array of Microtraps*, Nature **404**, 579 (2000).

[22] U. Tanaka, K. Suzuki, Y. Ibaraki, and S. Urabe, *Design of a Surface Electrode Trap for Parallel Ion Strings*, J. Phys. B At. Mol. Opt. Phys. **47**, (2014).

[23] M. Kumph, P. Holz, K. Langer, M. Meraner, M. Niedermayr, M. Brownnutt, and R. Blatt, *Operation of a Planar-Electrode Ion-Trap Array with Adjustable RF Electrodes*, New J. Phys. **18**, (2016).

[24] J. J. García-Ripoll, P. Zoller, and J. I. Cirac, *Speed Optimized Two-Qubit Gates with Laser Coherent Control Techniques for Ion Trap Quantum Computing*, Phys. Rev. Lett. **91**, 2 (2003).

[25] J. D. Wong-Campos, S. A. Moses, K. G. Johnson, and C. Monroe, *Demonstration of Two-Atom Entanglement with Ultrafast Optical Pulses*, Phys. Rev. Lett. **119**, 1 (2017).

[26] V. M. Schäfer, C. J. Ballance, K. Thirumalai, L. J. Stephenson, T. G. Ballance, A. M. Steane, and D. M. Lucas, *Fast Quantum Logic Gates with Trapped-Ion Qubits*, Nature **555**, 75 (2018).

[27] C. D. B. Bentley, A. R. R. Carvalho, D. Kielpinski, and J. J. Hope, *Fast Gates for Ion Traps by Splitting Laser Pulses*, New J. Phys. **15**, (2013).

[28] L. M. Duan, *Scaling Ion Trap Quantum Computation through Fast Quantum Gates*, Phys. Rev. Lett. **93**, 1 (2004).

[29] A. K. Ratcliffe, R. L. Taylor, J. J. Hope, and A. R. R. Carvalho, *Scaling Trapped Ion Quantum Computers Using Fast Gates and Microtraps*, Phys. Rev. Lett. **120**, 1 (2018).

[30] E. P. G. Gale, Z. Mehdi, L. M. Oberg, A. K. Ratcliffe, S. A. Haine, and J. J. Hope, *Optimized Fast Gates for Quantum Computing with Trapped Ions*, Phys. Rev. A **101**, 52328 (2020).

[31] E. Torrontegui, D. Heinrich, M. I. Hussain, R. Blatt, and J. J. Garc\'ia-Ripoll, *Ultra-Fast Two-Qubit Ion Gate Using Sequences of Resonant Pulses*, New J. Phys. **22**, 103024 (2020).

[32] R. L. Taylor, C. D. B. Bentley, J. S. Pedernales, L. Lamata, E. Solano, A. R. R. Carvalho, and J. J. Hope, *A Study on Fast Gates for Large-Scale Quantum*

*Simulation with Trapped Ions*, Sci. Rep. **7**, 46197 (2017).

[33] C. D. B. Bentley, A. R. R. Carvalho, and J. J. Hope, *Trapped Ion Scaling with Pulsed Fast Gates*, New J. Phys. **17**, (2015).

[34] D. Leibfried, B. DeMarco, V. Meyer, D. Lucas, M. Barrett, J. Britton, W. M. Itano, B. Jelenković, C. Langer, T. Rosenband, and D. J. Wineland, *Experimental Demonstration of a Robust, High-Fidelity Geometric Two Ion-Qubit Phase Gate*, Nature **422**, 412 (2003).

[35] P. L. Kapitza and P. A. M. Dirac, *The Reflection of Electrons from Standing Light Waves*, Math. Proc. Cambridge Philos. Soc. **29**, 297 (1933).

[36] J. A. Mizrahi, *Ultrafast Control of Spin and Motion in Trapped Ions*, Thesis 100 (2013).

[37] W. C. Campbell, J. Mizrahi, Q. Quraishi, C. Senko, D. Hayes, D. Hucul, D. N. Matsukevich, P. Maunz, and C. Monroe, *Ultrafast Gates for Single Atomic Qubits*, Phys. Rev. Lett. **105**, 1 (2010).

[38] A. K. Ratcliffe, L. M. Oberg, and J. J. Hope, *Micromotion-Enhanced Fast Entangling Gates for Trapped-Ion Quantum Computing*, Phys. Rev. A **101**, (2020)